\documentclass{ws-procs975x65}

\def\be {\begin{equation}}
\def\ee {\end{equation}}
\def\bea {\begin{eqnarray}}
\def\eea {\end{eqnarray}}
\def\ba {\begin{align}}
\def\ea {\end{align}}

\def\nn{\nonumber \\}

\begin{document}

\title{Testing the concordance model in cosmology with model-independent methods: some issues}
\author{Diego S\'aez-G\'omez}

\address{Departamento de F\'isica \& Instituto de Astrof\'isica e Ci\^encias do Espa\c{c}o,\\ 
Faculdade de Ci\^encias da Universidade de Lisboa, Edif\'icio C8, Campo Grande, P-1749-016
Lisbon, Portugal}

%

\begin{abstract}
Since the number of dark energy models have rapidly increased over the last years, some model-independent methods have been developed in order to analyse the cosmological evolution in a phenomenological way. In this manuscript, we analyse some of these approaches and their shortcomings to provide reliable information.
\end{abstract}
\bodymatter
\section{Introduction}
Since 1998 when a deviation on the luminosity distance of Supernovae Ia (Sne Ia) was observed by two independent groups \cite{SN1}, and later on by other proofs, the expansion of the universe is thought to be accelerating, a phenomena that has been widely accepted by the scientific community since then. In order to explain such behaviour of the universe expansion, plenty of theoretical models  have been proposed to sort out this challenge, under the name of dark energy. The list of theoretical models includes canonical/phantom scalar fields, vector fields, modifications of General Relativity (GR), or a cosmological constant, among others\cite{Nojiri:2010wj}. The latter, known as $\Lambda$CDM model , has been the one that serves as a reference to test the others.\\
However, plenty of models of dark energy are able to fit the data as good as $\Lambda$CDM model, what increases the difficulty to sort out this problem. This degeneracy among models does not seem to break at the background level at least with the present data, but other tests are required, as the analysis of the perturbations. Then, a useful tool can be to develop an approach able to provide a particular dynamical behaviour of dark energy without providing explicitly the underlying theoretical framework. Roughly speaking, an independent-model approach that serves just for phenomenological purposes. In this sense, some interesting approaches have been analysed over the last years. In this sense, some parametrizations of the dark energy equation of state (EoS) may provide  information about the dynamical evolution of dark energy, while other model-independent methods as cosmography may lead to information about the background evolution and the possibility of testing $\Lambda$CDM model. Here some issues regarding these model-independent methods are analysed and their usefulness for testing the concordance model of cosmology.
\section{Parametrizations of the equation of state for dark energy}
\label{reconstruction}
In order to test the dynamics of dark energy, some parametrizations of the EoS for dark energy have been proposed in the literature, in such a way that may show up how the evolution of dark energy is, at least at low redshifts, regardless of the underlying theory. In this sense, we may highlight the ones proposed  by Huterer and Turner in 2001\cite{Huterer:2000mj}, and the Chevallier-Polarski-Linder parametrization\cite{Huterer:2000mj}, which are given by
\be
w_{HT}(z) = w_0 + w_1 z, \quad w_{CPL}=w_0+w_1\frac{z}{1+z}\ .
\label{ST11}
\ee
The first parametrization, also called Linear Redshift Parametrization,  is assumed to describe the behaviour of the dark energy fluid at low redshift ($z < 1$), while the second one tends to a constant for large redshifts. Fitting both parametrizations with Supernovae Ia data, the best fit gives $w_0=-1.4$ and $w_1=1.67$ (HT parametrization), and $w_0=-0.82$ and $w_1=0.58$ (CPL parametrization), leading to a very similar goodness of fit in both cases. Note that both parametrizations include $\Lambda$CDM as a particular case which does not coincide with the best fit although can not be discarded. In addition, both phenomenological descriptions predict a very different EoS at $z=0$, where $w_{HT}=-1.4$ (phantom-like fluid) and $w_{CPL}=-0.82$, a degeneracy problem that commonly arise when comparing theoretical models directly.\\
Similarly, other authors have investigated other parametrizations of the EoS where a fast transition -to a phantom epoch- may occur\cite{Leanizbarrutia:2014xta}:
\begin{equation}
w_{1}(z)=- 1+ w_{0}\left[ \tanh\left(z-z_0\right)-1\right]\ ,\quad w_2(z)=-1+ w_0 \tanh\left(z-z_0\right)\ .  
\label{2.1}
\end{equation}
Here $w_{0}$ and $z_0$ are free parameters, being $z_0$ the turning point of both functions along the cosmological evolution and $w_0$ the value of the EoS parameter when $z\leq z_0$. Note also that for $w_{0}=0$, $\Lambda$CDM is recovered in both cases, while the phantom transition may never occur for some particular ranges of the free parameters. Consequently, in the first parametrisation in (\ref{2.1}), a future singularity will occur in case that $w_0>0$ whereas  the expansion would be smooth when $w_0<0$. In the second one, the value of $\tilde{w}_2$ is above -1 for negative (positive) values of $w_0$ and $z_0 > -1$ ($z_0 < -1$), so there is no singularity. For positive (negative) values of $w_0$ and $z_0 > -1$ ($z_0 < -1$), the value of $\tilde{w}_2$ is below $-1$ and a singularity occurs.  Thus, depending on the free parameters, the above models may lead to some kind of future singularity. Then, by using Sne Ia data, the best fits are obtained for both models in (\ref{2.1}) and compared with $\Lambda$CDM. The results are summarised in the following table, where the best $\chi^2$ is included as well as the reduced $\chi^2_{red}$ in order to compare the goodness of the fit for every model. As shown, every model gives a very similar fit, but different cosmological evolutions. Similarly, by using other data sets as BAO, the degeneracy of the results remains, such that parameterisations of the EoS for dark energy are not very useful to find out how the dynamics of dark energy are, at least with the available data sets.
\begin{table}[h!]  
\begin{center}
\begin{tabular}{ccccccc}
\hline
\hline
\bf{Model} & $\bf{\chi_{min}^2}$ &  $\bf{w_0}$ & $\bf{z_0}$ & $\bf{\Omega_{0m}}$ & $\bf{\chi_{red}^2}$\\
\hline \vspace{-5pt}\\
$\Lambda$CDM  & $542.685$ & - & - & $0.27 \pm 0.02$ & $0.978 $\\
\\
$ w_1(z)$ & $542.683$ & $0.0045 \pm 0.1$ & $-25 \pm 30$ & $0.27$ & $0.981 $ \\
\\
$ w_2(z)$ & $541.583$ & $-0.03 \pm 0.07$ & $22 \pm 45$ & $0.27$ & $0.979 $\\\\
 \hline \hline
\end{tabular}
\caption{Best fit for the models (\ref{2.1}) with $\Omega_{0m}=0.27$ by using the Sne Ia dataset \cite{union2.1}. The result for the $\Lambda$CDM model is also shown. \label{table1}}
\end{center}
\end{table}
\section{Cosmography}
\label{cosmography}
Another model-independent approach to test cosmology is the so-called cosmography, which is based solely on the cosmological principle regardless of the underlying theoretical model. To do so, the Hubble parameter is expanded in terms of an auxiliary variable\cite{Weinberg-Harrison}:
\be
H(z)=\frac{\dot{a}}{a}=H_0+H_{z0}z+\frac{H_{zz0}}{2}z^2+\frac{H_{zzz0}}{6}z^3+...\ ,
\label{cosmo1}
\ee
Here we have used the redshift $1+z=\frac{1}{a}$ as the auxiliary variable and the subscript $0$ refers to quantities evaluated today. Then, the cosmographic parameters are defined in terms of the derivatives of the Hubble parameter, or equivalently in terms of the scale factor as
\begin{equation}
H_0=\frac{\dot{a}_0}{a_0}\ ,\, q_0=-\frac{\ddot{a}_0}{a_0H^2_0}\ , \, j_0=\frac{a_0^{(3)}}{a_0H_0^3}\ , \,  s_0=\frac{a_0^{(4)}}{a_0H_0^4}\ ,  ...\ 
\label{cosmo2}
\end{equation}
where the dots are cosmic time derivatives. Then, by inserting (\ref{cosmo2}) in (\ref{cosmo1}), the Hubble parameter is written in terms of the cosmographic parameters which should be set with  observational data. However, note that the series (\ref{cosmo1}) converges for $|z|<1$. Hence, alternatively the expansion (\ref{cosmo1})  may be expressed in terms of another auxiliary variable $y = \frac{z}{1+z}$ which successes to describe the entire universe history by ensuring the convergence of the series\cite{Visser}. By generating mock data from a fiducial spatially flat $\Lambda$CDM model, where we have assumed the same redshifts as the Union 2.1 catalogue \cite{union2.1}, with errors of magnitude $\sigma_{\mu}=0.15$, we can show which variable and order of the series behaves better. We have run 100 simulations and fit the cosmographic parameters by using two different sets of parameters for each variable: $\bm{\theta_1} = \{H_0,q_0,j_0,s_0\}$ and $\bm{\theta_2} = \{H_0,q_0,j_0,s_0,l_0\}$, where $H_0$ is marginalised. Then, by using Monte Carlo Markov Chain (MCMC), the corresponding constraints for each set and each variable are obtained and shown in the following table. The table contains the number of times the true parameters were inside the confidence region bounds. It is expected the true value to lie within $1\sigma$, 
68\% of the times and $2\sigma$, 95\% of the times. The variable $z$ gives well-behaved coverage 
results for the set $\theta_1$, overestimates the errors while considering a higher order in the expansion $\theta_2$. On the other hand, the $y$-parametrisation gives completely biased 
estimators for the set $\theta_1$, and overestimates the errors for $\theta_2$. These results show that the variable $z$ is preferable in comparison with $y$ for testing models\cite{Busti:2015xqa}.
\begin{table}[htbp]
\caption{Coverage test for $\bm{\theta_1}$ and $\bm{\theta_2}$. Refer to the bulk of the text for further details.}
\label{tables1}
\begin{center}
\begin{tabular}{@{}cccccccccccccc@{}}
\hline
&     &     &     &  $\bm{\theta_1}$   &      &     &     &      &     &  $\bm{\theta_2}$   &      &     \\ \hline 
& \vline    &   $y$   &   \vline  & \vline    &  $z$   &   \vline &   & \vline    &   $y$   &   \vline  & \vline    &   $z$   &   \vline  \\
\hline  & $1\sigma$ & $2\sigma$ &
 3$\sigma$ & $1\sigma$ & $2\sigma$ &
 3$\sigma$ &  & $1\sigma$ & $2\sigma$ &
 3$\sigma$ & $1\sigma$ & $2\sigma$ &
 3$\sigma$
\\ \hline
$q_0$ & 26 & 32 & 42 & 67   & 27 & 6 &  & 82 & 12 & 6 & 82   & 18 & 0  \\ 
$j_0$ & 10  &  45 & 45 & 64  & 29 & 7 &  & 93  &  5 & 2 & 88  & 12 & 0  \\ 
$s_0$ & 10  & 67 & 23 & 83  & 15 & 2 &  & 92  & 7 & 1 & 93  & 6 & 1 \\
$l_0$ & - & - & - & - & - & - &  & 100 & 0 & 0 & 100 & 0 & 0 \\
\hline
\end{tabular}
\end{center}
\end{table}
\subsection{Testing $\Lambda$CDM with cosmography}
\begin{figure*}
\begin{center}
\includegraphics[width=0.4\textwidth]{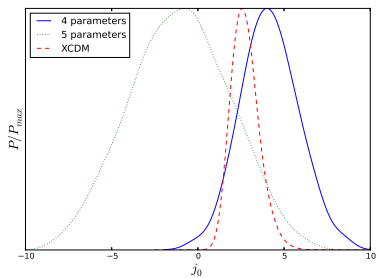}
\end{center}
\caption{Posterior probability for $j_0$ considering 4 parameters $(\bm{\theta_1})$, 5 parameters $(\bm{\theta_1})$ and XCDM model\cite{Busti:2015xqa}.}
\label{fig2}
\end{figure*}

Since flat $\Lambda$CDM model gives unequivocally $j_0=1$ regardless of the matter -and dark energy- content, cosmography can be used as a test for the $\Lambda$CDM model. In order to show the usefulness of the approach, we have generated Sne Ia data by assuming a XCDM model, different from $\Lambda$CDM, given by: $H^2/H_0^2=\Omega_m (1+z)^3+(1-\Omega_m)(1+z)^{3(1+w)}$ with $\Omega_m=0.3$ and $w=-1.3$, which gives $j_0=1.945$. Then, by assuming the variable $z$ and fitting the two parameters sets $\bm{\theta_1}$ (fourth order) and $\bm{\theta_2}$ (fifth order), we have tested whether the cosmographic approach can rule out $\Lambda$CDM by obtaining the posterior probability for $j_0$. As depicted in Fig.~\ref{fig2}, there is some evidence of $j_0 \neq 1$ when considering $\bm{\theta_1}$ but such evidence disappears when $\bm{\theta_2}$ is assumed. In addition, we have also included the posterior probability for $j_0$ while considering directly the expression of the Hubble parameter for the XCDM model, which leads to a much better fit.  Consequently, the constraints obtained by using cosmography present clear limits while comparing the concordance model with close-enough competitors. 
\subsection{Reconstructing dark energy models}
\begin{figure*}
\begin{center}
\includegraphics[width=0.3\textwidth]{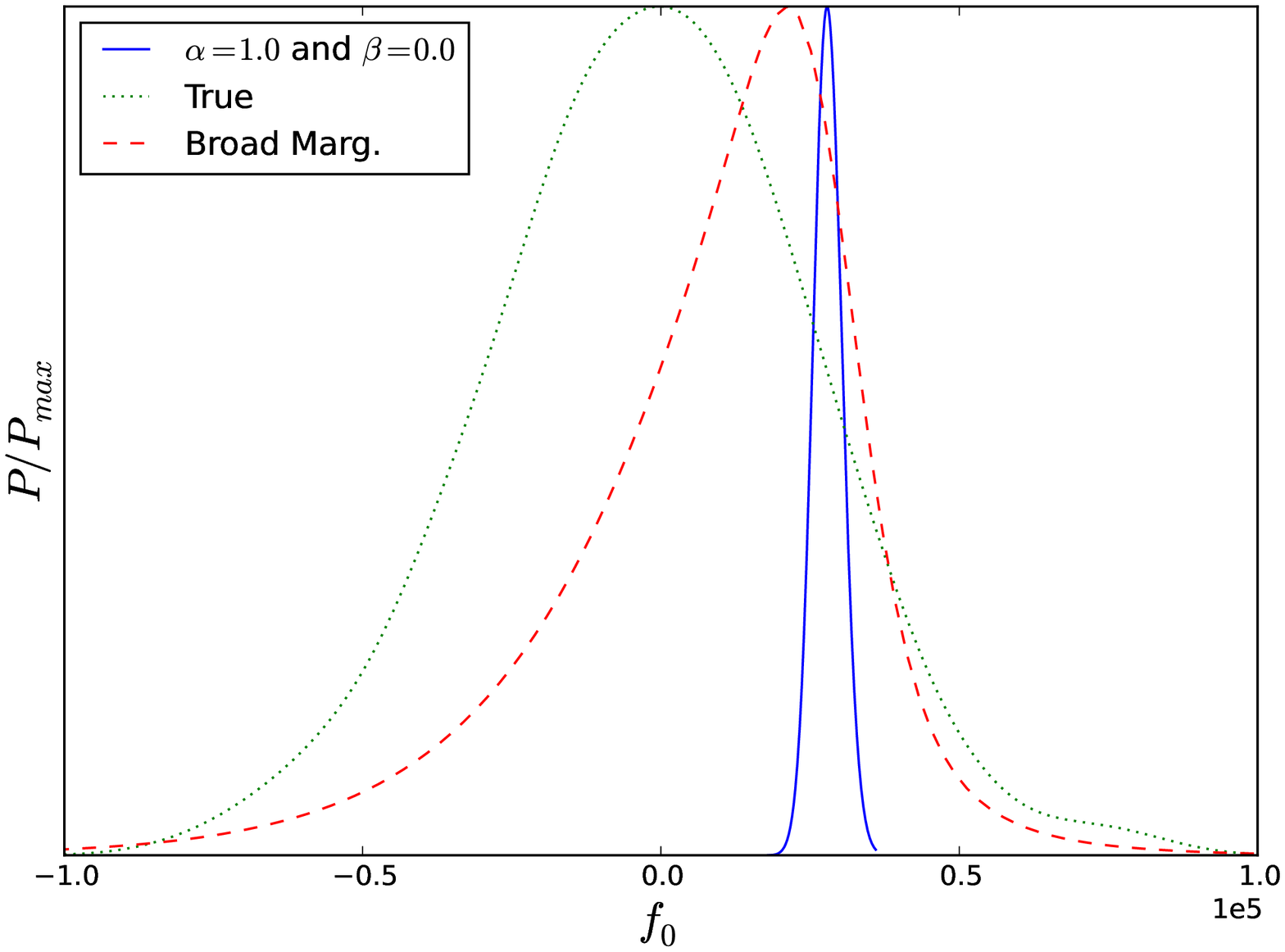}
\includegraphics[width=0.3\textwidth]{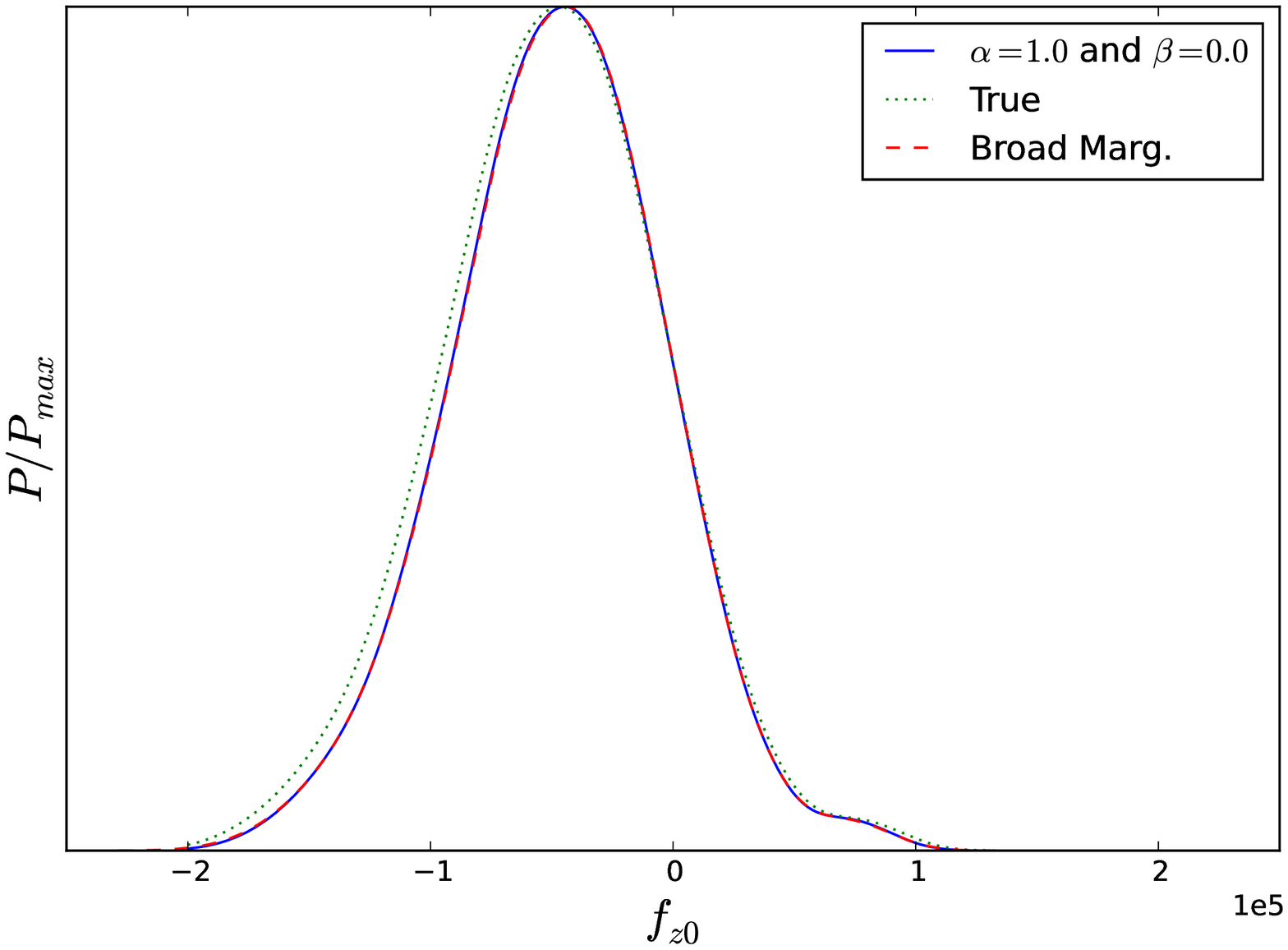}
\includegraphics[width=0.3\textwidth]{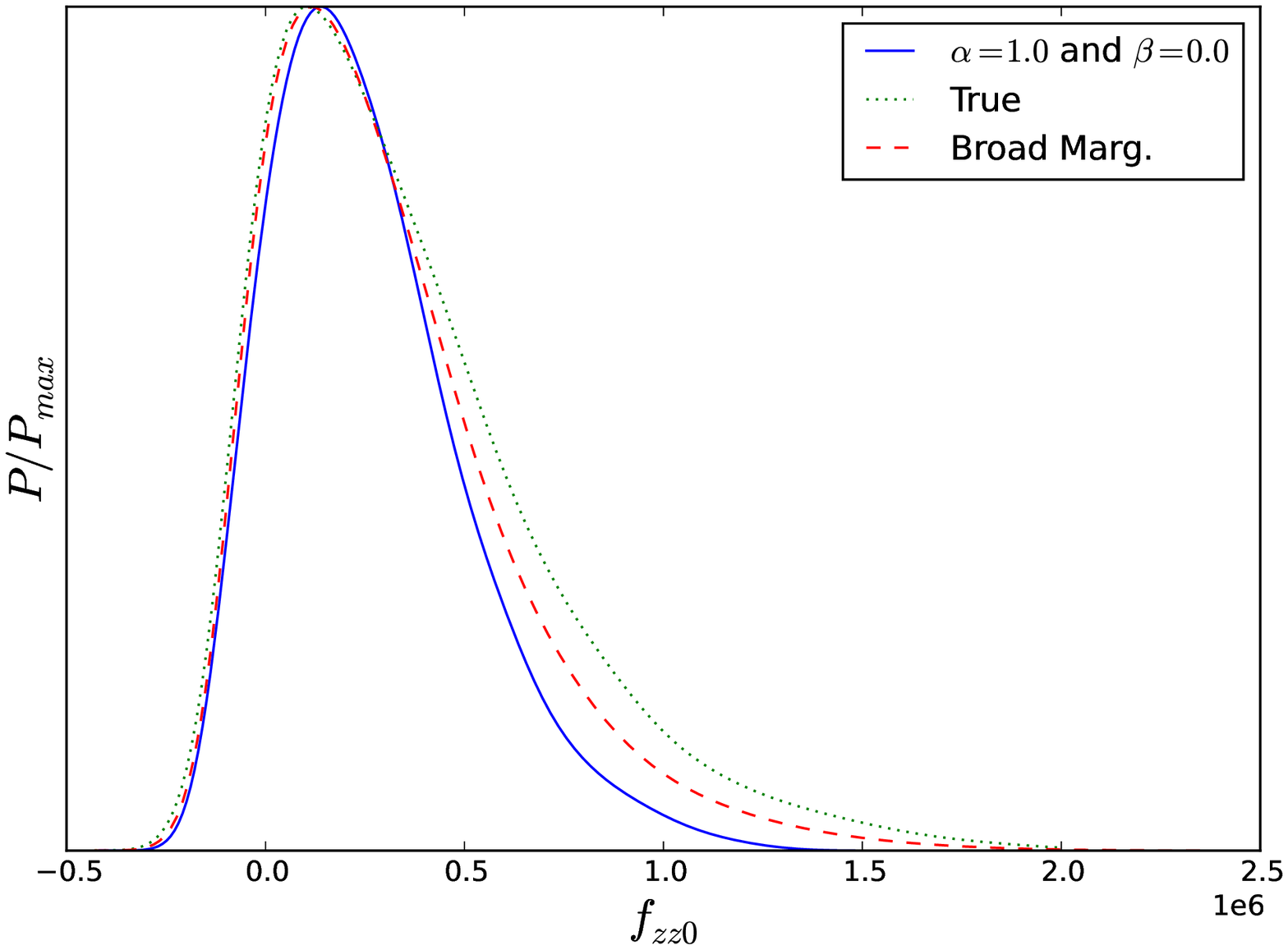}
\end{center}
\caption{Probabilities for $f(R)$ and its derivatives evaluated today and the effects of the different choices of the free parameters $\alpha$ and $\beta$\cite{Busti:2015xqa}.}
\label{fig3}
\end{figure*}

Cosmography have been also used for reconstructing some particular models for dark energy, since the underlying action can be expanded around $z=0$ and a correspondence with the cosmographic parameters is obtained. For instance, by considering the usual quintessence/phantom scalar field model for dark energy, $\mathcal{S}=\int {\rm d}^4x^{}\sqrt{-g}\left[-\frac{1}{2} \omega (\phi)\partial_{\mu} \phi \partial^{\mu }\phi -V(\phi )\right]\ ,$ where $\omega(\phi)$ is the factor that renormalises the scalar field $\phi$ and $V(\phi)$ 
its potential. Then, the derivatives of the potential evaluated today, i.e., at redshift zero,  can be expressed in terms of the cosmographic parameters as follows
\bea
\label{DE3}
&\frac{V_0}{H_0^2}&=2-q_0-\frac{3\Omega_m}{2}\ , \quad \frac{V_{z0}}{H_0^2}=4+3q_0-j_0-\frac{9\Omega_m}{2}\ , \nn
&\frac{V_{2z0}}{H_0^2}&=4+8q_0+j_0(4+q_0)+s_0-9\Omega_m\ ,\nn
\eea
where we have used the FLRW equations for this model. We can assume  $\Omega_m \approx 2/3(1+q_0)$ by considering the universe to be close enough to $\Lambda$CDM today, what yields a one-to-one correspondence between the derivatives of the potential and the cosmographic parameters. Then, fitting the cosmographic parameters leads to constraints over the derivatives of the scalar potential. However, as shown by Dunsby et al (2015)\cite{Busti:2015xqa}, the constraints obtained by means of the cosmographic approach provides larger errors than other model-independent methods.\\
Similarly, higher-order derivatives models as Galileons or $f(R)$ gravities can be evaluated at $z=0$ in terms of the cosmographic parameters. However, in this case the higher number of degrees of freedom does not allow to get a one-to-one correspondence as in (\ref{DE3}). In order to show this, let us consider $f(R)$ gravity, whose derivatives evaluated today lead to
\begin{eqnarray}
\label{DE4}
&&\frac{f_0}{6H_0^2}\,=\,-\alpha q_0 +\Omega_m + 6\beta\left( 2+q_0-j_0\right)\ , \quad \frac{f_{z0}}{6H_0^2}\,=\,\alpha\left(2+q_0-j_0\right)\ , \label{f2z0} \\
&&\frac{f_{2z0}}{6H_0^2}\,=\,6\beta\left(2+q_0-j_0\right)^2+\alpha\left[2+4q_0+(2+q_0)j_0+s_0\right]\,.\nonumber 
\end{eqnarray}
In this case, there are two extra free parameters, $\frac{{\rm d}f}{{\rm d}R}|_{R=R_0}=\alpha$ and $\frac{{\rm d}^2f}{{\rm d}R^2}|_{R=R_0}=\frac{\beta}{H_0^2}$.
This means we need either theoretical priors or additional tests since data does not provide any constraints over $\alpha$ and $\beta$. Some previous works assumed the values of $\alpha=1$ and $\beta=0$ a priori, such that the model coincides with General Relativity at $z=0$, but this may lead to instabilities. Let us illustrate the difficulties in getting good constraints for $f(R)$ gravities by generating mock data and assuming some sensible priors over the aforementioned parameters for the following toy-model: $f(R)=R+aR^2+bR^3$ where $\alpha=2.81$ and $\beta=0.06$. In Fig.~\ref{fig3}, the probability for $\{f_0, f_{z0}, f_{zz0}\}$  are depicted. Here three different hypotheses have been assumed: the true values of $\{\alpha, \beta\}$, $\{\alpha=1, \beta=0\}$ and a ``broad'' marginalisation ($\alpha \sim N(1,0.05)$ and $\beta \sim N(0.07,0.05)$). The probability of $f_0$ is highly dependent on the choice of $\{\alpha, \beta\}$ which may even lead to ruling out the true values of $f_0$, whether are not known in advance - as is the case when dealing with real data. There are not large differences for the values of $f_{z0}$ and $f_{zz0}$, but the errors are so large than almost any $f(R)$ may be valid, leading to a completely degenerated result. Consequently cosmography is extremely weak when reconstructing $f(R)$ gravities, since it does not allow to distinguish among different Lagrangians.
\section*{Acknowledgments}
D.S.-G. acknowledges support from a postdoctoral fellowship Ref.~SFRH/BPD/95939/2013 by Funda\c{c}\~ao para a Ci\^encia e a Tecnologia (FCT, Portugal) and the support through the research grant UID/FIS/04434/2013 (FCT, Portugal).

\end{document}